\documentclass[12pt]{article}

\usepackage{amsmath,amssymb,amsthm}

\usepackage{xchicago}
\makeatletter

\def\citeAPY{\def\@citeseppen{-1000}%
    \def\@cite##1##2{##1\if@tempswa , ##2\fi}%
    \def\citeauthoryear##1##2##3{##2 (##3)}\@internalcite}

\makeatother
\newcommand {\citeAY} [1] {\citeNP {#1}}%
\renewcommand {\showoriginalref}[1]{}
\renewcommand {\showCODEN}[1]{}
\renewcommand {\showISSN}[1]{}
\renewcommand {\showMR}[3]{}

\newcommand\eq[1] {(\ref{#1})}

\newcommand\fig[1] {\ref{fig:#1}}

\newcommand\labfig[1] {\label{fig:#1}}

\newcommand{\beqa}{\begin{eqnarray}}
\newcommand{\eeqa}[1]{\label{#1}\end{eqnarray}}
\newcommand{\beq}{\begin{equation}}
\newcommand{\eeq}[1]{\label{#1}\end{equation}}

\newcommand{\Gl}{\lambda}

\def\Bb{{\bf b}}

\def\Bu{{\bf u}}
\def\Bv{{\bf v}}

\def\Bx{{\bf x}}

\def\BA{{\bf A}}
\def\BB{{\bf B}}
\def\BC{{\bf C}}

\def\BF{{\bf F}}

\def\BI{{\bf I}}

\def\BM{{\bf M}}

\def\BR{{\bf R}}

\def \ba {\begin{array}}
\def \ea {\end{array}}

\def \refe #1.{(\ref{#1})}
\def \reff #1.{figure~\ref{#1}}
\def \refs #1.{section~\ref{#1}}
\def \refss #1.{subsection~\ref{#1}}
\def \refD #1.{Definition~\ref{#1}}
\def \refT #1.{Theorem~\ref{#1}}
\def \refL #1.{Lemma~\ref{#1}}
\def \refC #1.{Corollary~\ref{#1}}
\def \refP #1.{Proposition~\ref{#1}}
\def \refR #1.{Remark~\ref{#1}}
\def \refE #1.{Example~\ref{#1}}
\def \refN #1.{Notation~\ref{#1}}

\usepackage{graphics}
\usepackage{amssymb}
\begin{document}
\vspace{-1in}
\title{Adaptable nonlinear bimode metamaterials using rigid bars, pivots, and actuators.}
\author{Graeme Walter Milton\\
\small{Department of Mathematics, University of Utah, Salt Lake City UT 84112, USA}}
\date{}
\maketitle
\begin{abstract}
A large family of periodic planar non-linear bimode metamaterials are constructed from rigid bars and pivots. They are affine materials in 
the sense that their macroscopic deformations are only affine deformations: at large distances any deformation must be close to an affine
deformation.  Bimode means that the paths of all
possible deformations of Bravais lattices that preserve the periodicity of the lattice 
lie on a two dimensional surface in the three dimensional space of invariants
describing the deformation (excluding translations and rotations). By adding two actuators inside a single microscopic cell
one can control the macroscopic deformation, which may be useful for the design of adaptive structures. The design
of adaptable nonlinear affine trimode metamaterials (for which the macroscopic deformations lie within a three-dimensional
region in the space of invariants) is discussed, although their realization remains an open problem. Examples are
given of non-affine unimode and non-affine bimode materials. In these materials the deformation can vary from cell to cell
in such a way that the macroscopic deformation is non-affine. 
\end{abstract}
\vskip2mm
\noindent Keywords: A. microstructures, B. inhomogeneous material; B rods and cables; B foam material; bimode material  
\noindent 
\vskip2mm
\section{Introduction}
\setcounter{equation}{0} 
Here we consider the deformation of planar periodic lattices of rigid (inextensible) bars with pivot joints. 
Since they are periodic they have a two dimensional Bravais lattice consisting of points 
\beq \Bx=i\Bu+j\Bv, \eeq{1.1}
as $i$ and $j$ range over all integers, where $\Bu$ and $\Bv$ are the lattice vectors. Let $\BF$ 
be the deformation matrix with the vectors $\Bu$ and $\Bv$ as columns. 
Then under a rotation $\BR$, since $\Bu$ and $\Bv$ transform
to $\BR\Bu$ and $\BR\Bv$ it follows that $\BF$ transforms to $\BR\BF$ leaving 
the symmetric Cauchy-Green matrix $\BA=\BF^T\BF$ invariant. There are of course many other Bravais lattices. In particular, 
there are lattices with lattice vectors
\beq \Bu'=k\Bu+\ell\Bv, \quad \Bv'=m\Bu+n\Bv, \eeq{1.2}
for any choice of the integers $k$, $\ell$, $m$ and $n$ such that $\Bu'$ and $\Bv'$ are independent. 
The corresponding deformation matrix $\BF'$, which has $\Bu'$ and $\Bv'$ as columns, is given by
\beq \BF'=\BF\BM,\quad {\rm where}~
\BM=\begin{pmatrix} k & m \\ \ell & n \end{pmatrix}, 
\eeq{1.3}
and $\BM$ is non-singular. Other Bravais lattices will correspond to matrices $\BM$ with elements that are not necessarily integers.
As the material continuously deforms, with the Bravais lattice having lattice vectors $\Bu$ and $\Bv$ 
undergoing an affine transformation, $\BF$ follows some trajectory $\BF(t)$ beginning at $t=t_0$. Associated with the deformation
is a path
\beq \BC(t)=[\BF(t_0)^T]^{-1}\BF(t)^T\BF(t)[\BF(t_0)]^{-1} \eeq{1.4}
in the three dimensional space of symmetric matrices beginning at $\BC(t_0)=\BI$.
If we chose a different Bravais, such as that in \eq{1.2}, this path, for the same deformation, would be the same since
\beq [\BM^T\BF(t_0)^T]^{-1}\BM^T\BF(t)^T\BF(t)\BM[\BF(t_0)\BM]^{-1}=[\BF(t_0)^T]^{-1}\BF(t)^T\BF(t)[\BF(t_0)]^{-1}. \eeq{1.5}
On the other hand there can be continuous deformations in which the larger lattice undergoes an affine transformation, but the
smaller lattice is distorted. So it is possible that there exist additional paths of deformation $\BC(t)$ for the larger lattice which
are not accessible to the smaller lattice, see figure \fig{1} and the examples in  \citeAPY{Kapko:2009:CLI}, for instance. 

A material is classed as null-mode if the
only possible continuous deformation path for any Bravais lattice is the point $\BC(t)=\BI$. Some internal motions could be possible but
macroscopically the material is rigid to periodic deformations (in the sense that for any Bravais lattice the only continuous deformations that keep
the periodicity of that lattice only rotate and translate the lattice). A material is classed as unimode if it is not null-mode and the possible deformations
$\BC(t)$ for any Bravais lattice all lie on the same one dimensional curve. A material is classed as bimode if it is not null-mode or unimode 
and the possible deformations $\BC(t)$ for any Bravais lattice all lie on the same two dimensional surface. A material is classed as trimode
if it is not null-mode, unimode, or bimode. These definitions extend the definitions of unimode, bimode and trimode introduced by 
\citeAPY{Milton:1995:WET} in the context of planar linear elasticity (see also chapter 30 of \citeAY{Milton:2002:TOC}). The definitions
extend to three dimensions in the obvious way, and in three dimensions one can also have quadramode, pentamode, and hexamode 
nonlinear materials (\citeAY{Milton:2012:CCM}).
Recently, a class of linearly elastic pentamodes suggested by \citeAPY{Milton:1995:WET} were experimentally realized by  
\citeAPY{Kadic:2012:PPM}.

Periodic materials made from bars and pivots can be further classed into ``affine materials''
for which an affine transformation is the only possible macroscopic deformation, and non-affine materials like that in figure \fig{1},
which can have other macroscopic deformations. In these non-affine materials the deformation varies from cell to cell in such a way that the
macroscopic deformation is non-affine. More precisely a material is affine if and only if any deformation $\Bx'(\Bx)$ (defined only for points
$\Bx$ on the rigid bars which get moved to $\Bx'(\Bx)$) necessarily approaches $\BB\Bx+\Bb$ as $|\Bx|\to\infty$, for some non-singular matrix $\BB$ and vector $\Bb$ (dependent on the deformation) 
to within terms of $o(|\Bx|)$. A material which isn't affine is non-affine. A  material is macroscopically rigid if and only if any deformation $\Bx'(\Bx)$
necessarily approaches $\BR\Bx+\Bb$ as $|\Bx|\to\infty$, for some rotation $\BR$ and vector $\Bb$, to within terms of $o(|\Bx|)$. We do not know if
a null-mode material is necessarily macroscopically rigid.  

There are lots of examples of affine unimode materials, see the companion paper (\citeAY{Milton:2012:CCM}) and references therein. Section 2
of this paper gives an example of a non-affine unimode material and an example of a non-affine bimode material. It also
shows in section 3 how a large family of affine bimode materials can be
constructed using bars and pivots. [Based on the figures of \citeAPY{Kapko:2009:CLI} the Roman mosaic net may be another example of an affine bimode
material, but we are not sure of this.]  It is easy to construct non-affine trimode  materials from bars and pivots: figure \fig{1} gives an example.
We do not have any examples of affine trimode materials, although in section 4 we suggest a possible route towards obtaining them.

Although the constructions are two-dimensional they can easily be extended into the third dimension, to obtain
three-dimensional materials, by replacing rigid bars with rigid sheets, and pivot joints by edge joints. Then these structures 
can then be replaced by trusses of rigid bars and pivots. In this transformation two dimensional $n$-mode materials become
three dimensional $n$-mode materials, and affineness, or non-affineness, is preserved.

\citeAPY{Guest:2003:DRS} raised the interesting question as to whether it is possible to build a 
periodic pin-jointed structure of rigid bars having no easy modes of deformation and such that replacing any bar by an actuator and changing its length leads to a 
change of the geometry of the structure, and not to self stress. 
They pointed out this is relevant to designing adaptive structures. They found that no such periodic pin-jointed structures of rigid bars can exist.
On the other hand, the affine bimode materials considered here have the property that 
they become macroscopically rigid if two (appropriately placed) bars 
are added to just one cell in the structure. Changing the lengths of these bars (within limits) leads only to a change of the geometry of the structure, and not to self stress. 
Replacing the bars by actuators leads to interesting adaptive structures. 

\begin{figure}[htbp]
\vspace{1.5in}
\hspace{0.0in}
{\resizebox{1.0in}{0.5in}
{\includegraphics[0in,0in][6in,3in]{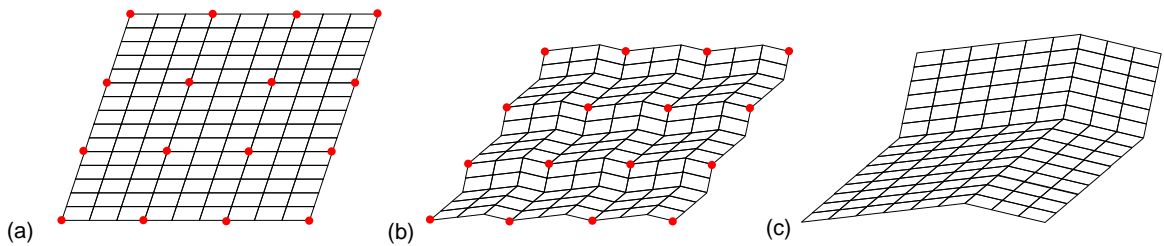}}}
\vspace{0.1in}
\caption{A parallelogram array of rigid bars can deform from (a) to (b). The red dots denote the Bravais lattice which undergoes
an affine transformation. As a result the array is not unimode, but instead trimode. It is a non-affine material as it has 
macroscopic deformations which are non-affine as in (c).}
\labfig{1}
\end{figure}

\section{Non-affine unimode and bimode materials}
\setcounter{equation}{0} 
To construct a non-affine unimode material we construct a rectangular box which remains rectangular as the microstructure inside the box
deforms, with two opposing sides remaining rigid. The key to the construction is Hart's A frame (\citeAY{Hart:1877:SCP}) 
which is a device
for obtaining exact straight line motion: see figure \fig{1.1}(a). The mechanism has the property that the lengths $AP$ and $BP$
remain equal as the structure deforms. Thus the point $P$ traces out a vertical straight line which is a segment of the perpendicular
bisector of $AB$. As a consequence if we add supports as in figure \fig{1.1}(b) we obtain a table which deforms as an isosceles trapezoid
$GHJK$. Attaching two such tables, which are $180^\circ$ rotations of each other, by their feet, as in figure \fig{1.1}(c), gives 
the unit cell of a unimode material such that $\Gl_2$ is free to vary (in an interval) 
but $\Gl_1$ remains fixed.

To construct a non-affine bimode material we construct a rectangular box which remains rectangular as the microstructure inside the box
deforms and is such that the two side lengths are free to vary independently within limits. This is done by
adding an additional table (with ``the feet extending beyond the table top''), and another table which is a $180^\circ$
rotation of it and placing them back to back with the structure of  figure \fig{1.1}(c) to give the unit cell structure of figure \fig{1.2}.

That these unimode and bimode structures are non-affine is easily seen from figure \fig{1.3}.

\begin{figure}[htbp]
\vspace{3.0in}
\hspace{0.5in}
{\resizebox{1.0in}{0.5in}
{\includegraphics[0in,0in][4in,2in]{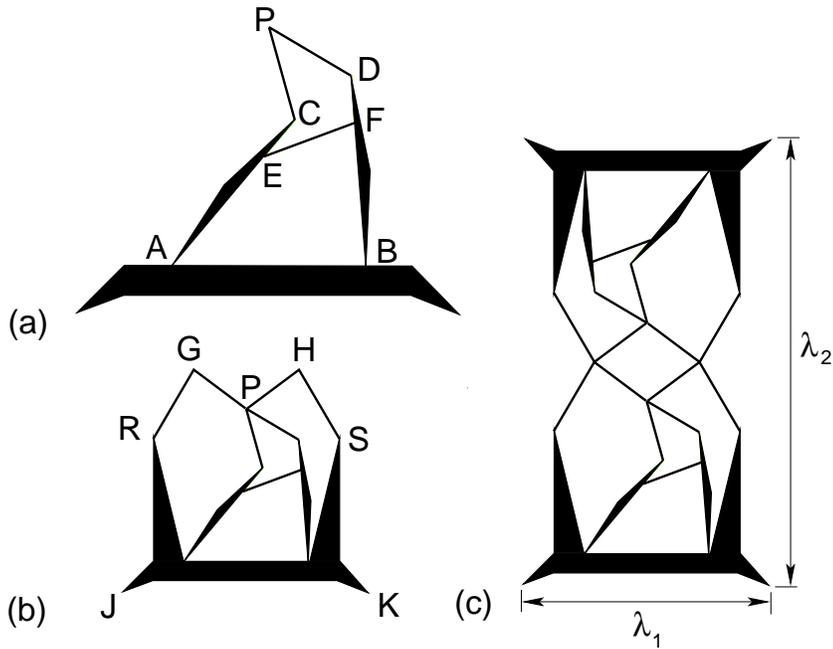}}}
\vspace{0.1in}
\caption{Figure (a) shows Hart's A frame. The black regions are rigid material. 
The lengths are such that $AC=BD=AB=a$, $PC=PD=EF=b$, and $CE=DF=c$ and
$b^2=ac$. For example, one may take $a=4$, $b=2$ and $c=1$. The triangle $APB$ remains isosceles as the structure deforms.
In (b) is a (inverted)``table'': we add two triangular pillars to the base of (a) and two
bipod supports with lengths $RG=SH$ and $GP=PH$ so that $GHJK$ remains
an isosceles trapezoid as the structure deforms, and $JK$ remains rigid. If we attach a table to an inverted table, with the 
same structure but rotated by $180^\circ$ as
in (c) we obtain the unit cell of the desired non-affine unimode material: $\Gl_2$ is free to vary (in an interval) but
$\Gl_1$ remains fixed.
}
\labfig{1.1}
\end{figure}

\begin{figure}[htbp]
\vspace{2.0in}
\hspace{0.5in}
{\resizebox{1.0in}{0.5in}
{\includegraphics[0in,0in][4in,2in]{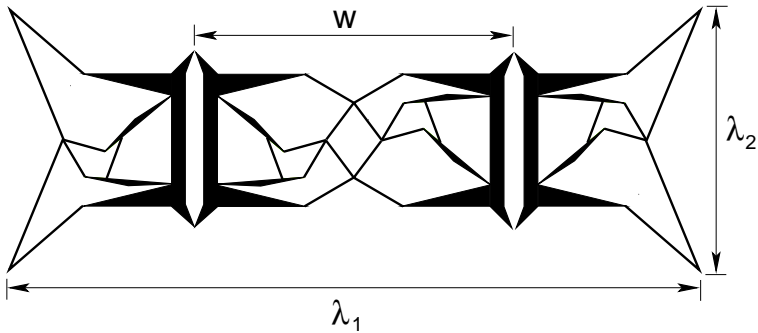}}}
\vspace{0.1in}
\caption{By combining four tables as shown here, we obtain the unit cell of the desired non-affine bimode material.
Note that we are free to adjust $\Gl_2$ (within limits) and given $\Gl_2$ we are still free to adjust $\Gl_1$ by varying $w$. }
\labfig{1.2}
\end{figure}

\begin{figure}[htbp]
\vspace{1.0in}
\hspace{0.5in}
{\resizebox{1.0in}{0.5in}
{\includegraphics[0in,0in][4in,2in]{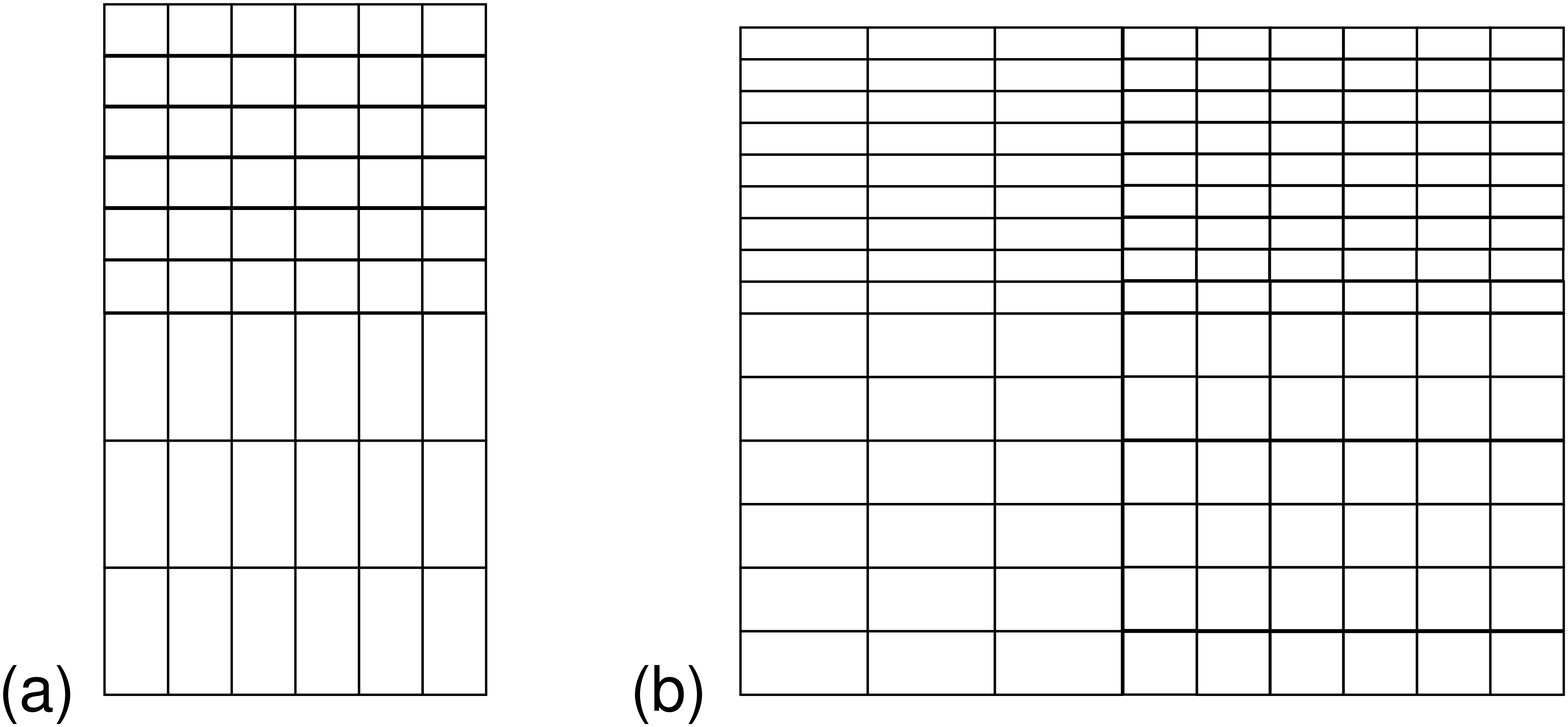}}}
\vspace{0.1in}
\caption{Figure (a) shows an exaggerated macroscopic deformation of a unimode material with each cell having the 
structure of figure \protect{\fig{1.1}}. Figure (b) shows an exaggerated macroscopic deformation of a 
bimode material with each cell having the structure of figure \protect{\fig{1.2}}. Therefore both are non-affine materials.}
\labfig{1.3}
\end{figure}

\section{Bimode materials with structure controlled by two actuators}
\setcounter{equation}{0} 

The first key component in the design of the desired bimode structures is what we call a u-structure (u for unimode), which is a triangle $ABC$
with one internal degree of freedom. An example is shown in figure \fig{2}.
A u-structure has three support points labeled $A$, $B$ and $C$, and is such that the lengths $AB$, $BC$ and $CA$ are not rigid but
are positive valued continuous functions $f(t)$, $g(t)$, and $h(t)$ of some variable $t$ which varies in some range $t_0^-\leq t \leq t_0^+$ 
as the structure deforms. Thus specification of one length determines the other two lengths if not uniquely, 
then within a finite (usually small) number of possibilities. The u-structure could be a triangle of unimode 
metamaterial, and from the results of \citeAPY{Milton:2012:CCM} it follows that $(f(t), g(t), h(t))$ can be made arbitrarily close to any desired
trajectory in three-dimensional space. It is helpful to assume the structure remains inside the triangle $ABC$ for
a some interval of values of $t$. Deformations with $t$ outside this interval may also be considered but then one runs
the risk of the structure colliding with other structures. 

\begin{figure}[htbp]
\vspace{2.5in}
\hspace{1.0in}
{\resizebox{2.0in}{1.0in}
{\includegraphics[0in,0in][7in,3.5in]{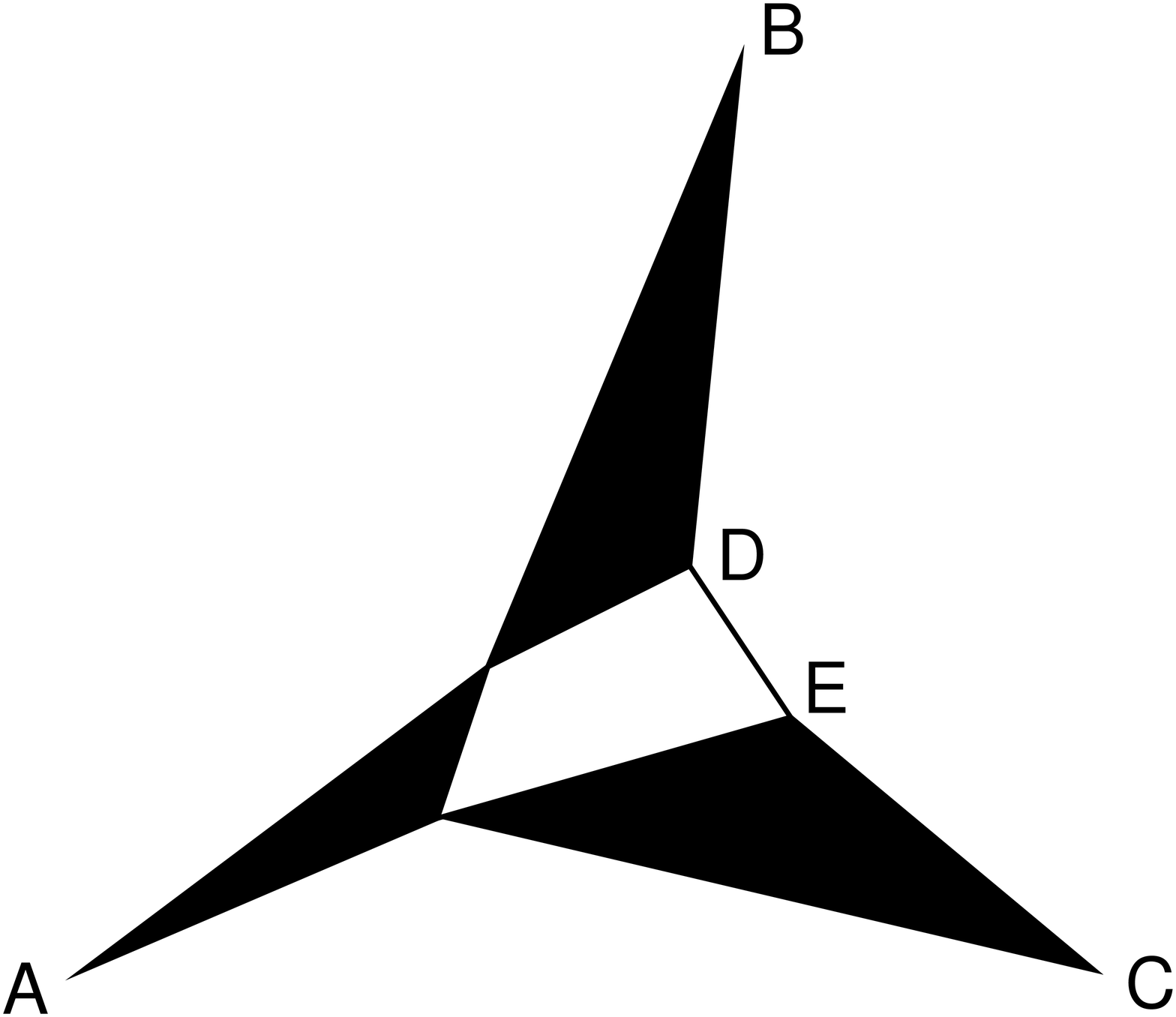}}}
\vspace{0.1in}
\caption{Example of a u-structure. The black triangles are rigid and $DE$ is a rigid bar.}
\labfig{2}
\end{figure}

The second key component in the design of the desired bimode structures is what we call a b-structure (b for bimode). It is a 
quadrilateral $ABCD$ with two internal degrees of freedom. It has sides with lengths $AB$, $BC$, $CD$ and $DA$ that are not rigid, 
but such that the lengths of any pair of adjoining
sides, such as $DA$ and $AB$ can range anywhere inside some open connected region of two-dimensional space (i.e. they are not
confined to lie on a curve) and that knowledge of these two lengths completely specifies the quadrilateral up to a finite
number of possibilities. In other words given the lengths $DA$ and $AB$ (or any other pair of adjoining lengths) one can
determine (up to a finite number of possibilities) the included angle $DAB$ and the lengths of the two remaining sides $BC$ and $CD$.
Additionally, the quadrilateral does not remain a trapezoid for arbitrary deformations of the structure, or if it does the lengths of the
two parallel sides are not slaves to each other but instead determine the shape of the trapezoid. 

Note that the quadrilateral of figure \fig{3}(a) is not a b-structure even though it has two internal degrees of freedom. The
pair of adjoining lengths $AB$ and $BC$ are confined to a curve and do not determine the other lengths $CD$ and $DA$. There
are continuous deformations of the structure, such as from  figure \fig{3}(a) to figure \fig{3}(b) that leave the triangle
$ABC$ unchanged, but not the quadrilateral $ABCD$. As a consequence, if we tile all space with these quadrilaterals and $180^\circ$ rotations of them, then
the resulting tiling can deform in many ways, such as illustrated in \fig{3}(c): the resulting structure is a non-affine trimode material
and has many internal degrees of freedom. 

\begin{figure}[htbp]
\vspace{7.0in}
\hspace{0.5in}
{\resizebox{2.0in}{1in}
{\includegraphics[0in,0in][8in,4in]{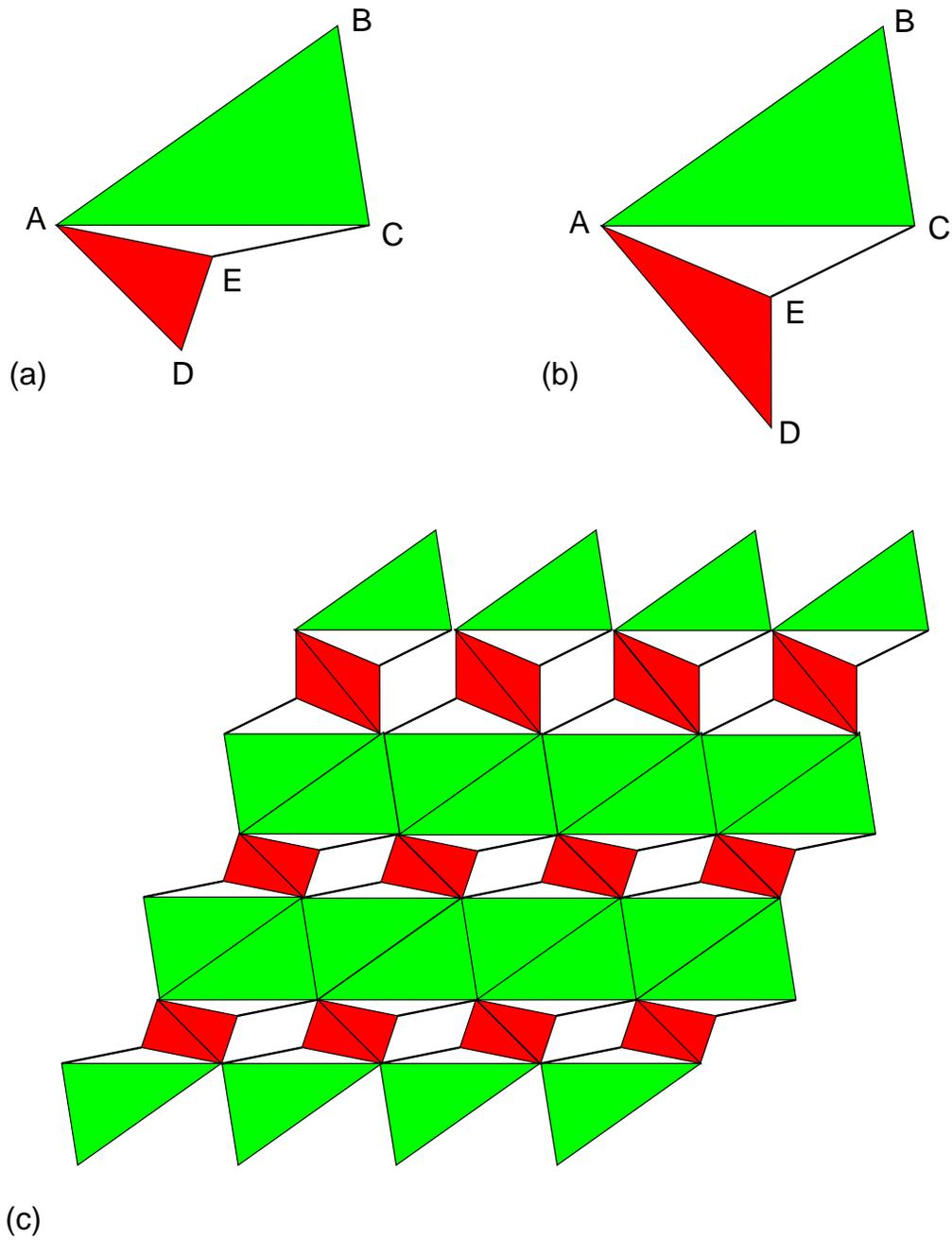}}}
\vspace{0.1in}
\caption{Figure (a) shows a quadrilateral $ABCD$ which is not a b-structure, but can deform as in (b) leaving the triangle $ABC$ unchanged. As a result the
tiling of space by these quadrilaterals can deform in many ways as illustrated, for example, in (c). Here the
green and red triangles are u-structures and $CE$ is a bar of fixed length}
\labfig{3}
\end{figure}

Neither is the trapezoid of figure \fig{3.5} a b-structure when the length $AB$ always remains twice (or some constant times)
the length $CD$, as the structure deforms. The structure is assumed to have two internal degrees of freedom, one being the length $CD$.
So even with the length $CD$ held constant the structure can deform from (a) to (b). A tiling of space 
by such trapezoids can deform in many ways as seen from figure \fig{3.5}(c)

\begin{figure}[htbp]
\vspace{3.0in}
\hspace{0in}
{\resizebox{3.0in}{1.5in}
{\includegraphics[0in,0in][8in,4in]{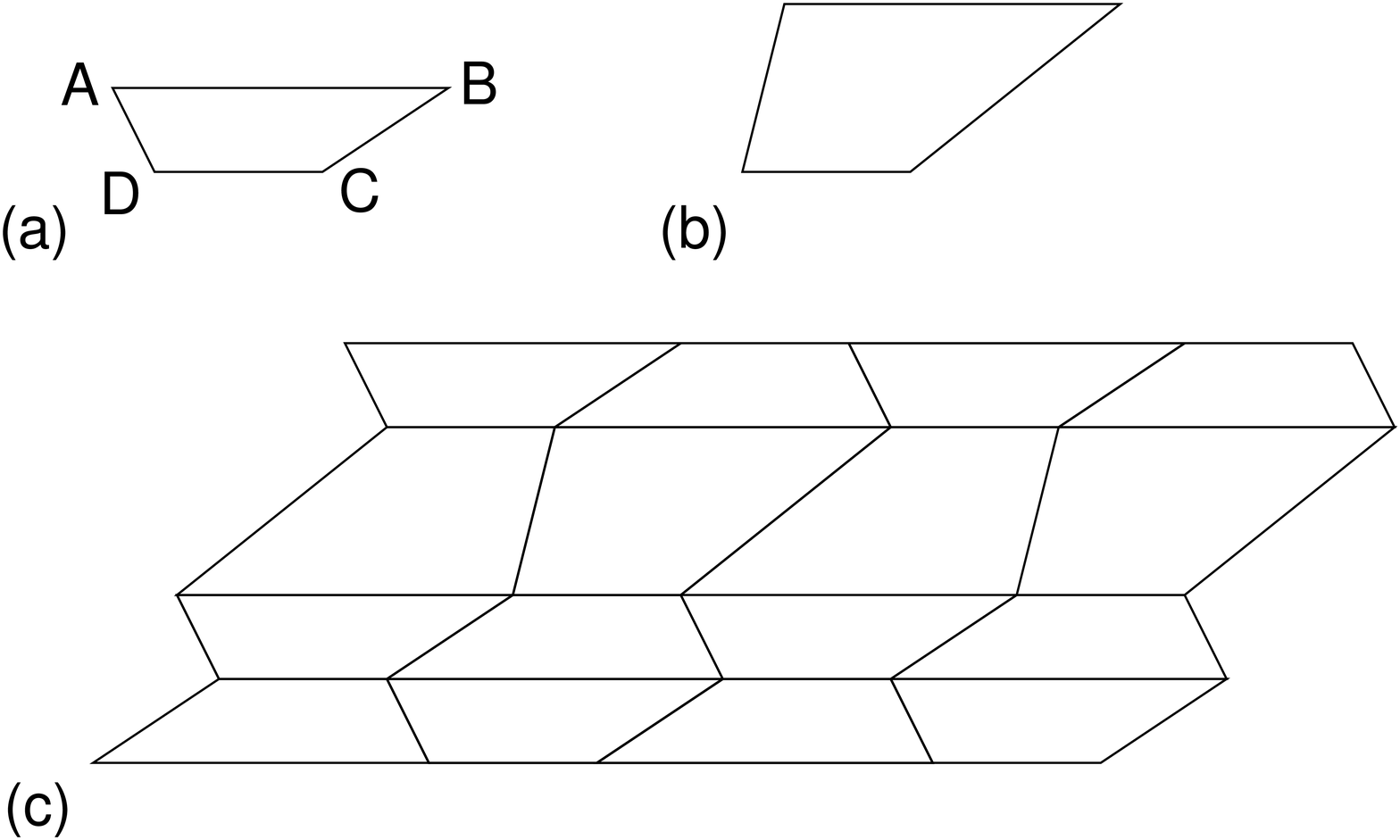}}}
\vspace{0.1in}
\caption{Figure (a) shows a trapezoid $ABCD$ which is not a b-structure when the length $AB$ always remains twice (or some constant times)
the length $CD$. If the length of $CD$ remains fixed the structure can deform from (a) to (b), and a lattice of such cells can deform in 
many ways as illustrated, for example, in (c).}
\labfig{3.5}
\end{figure}

\begin{figure}[htbp]
\vspace{4.0in}
\hspace{0.0in}
{\resizebox{2.0in}{1in}
{\includegraphics[0in,0in][8in,4in]{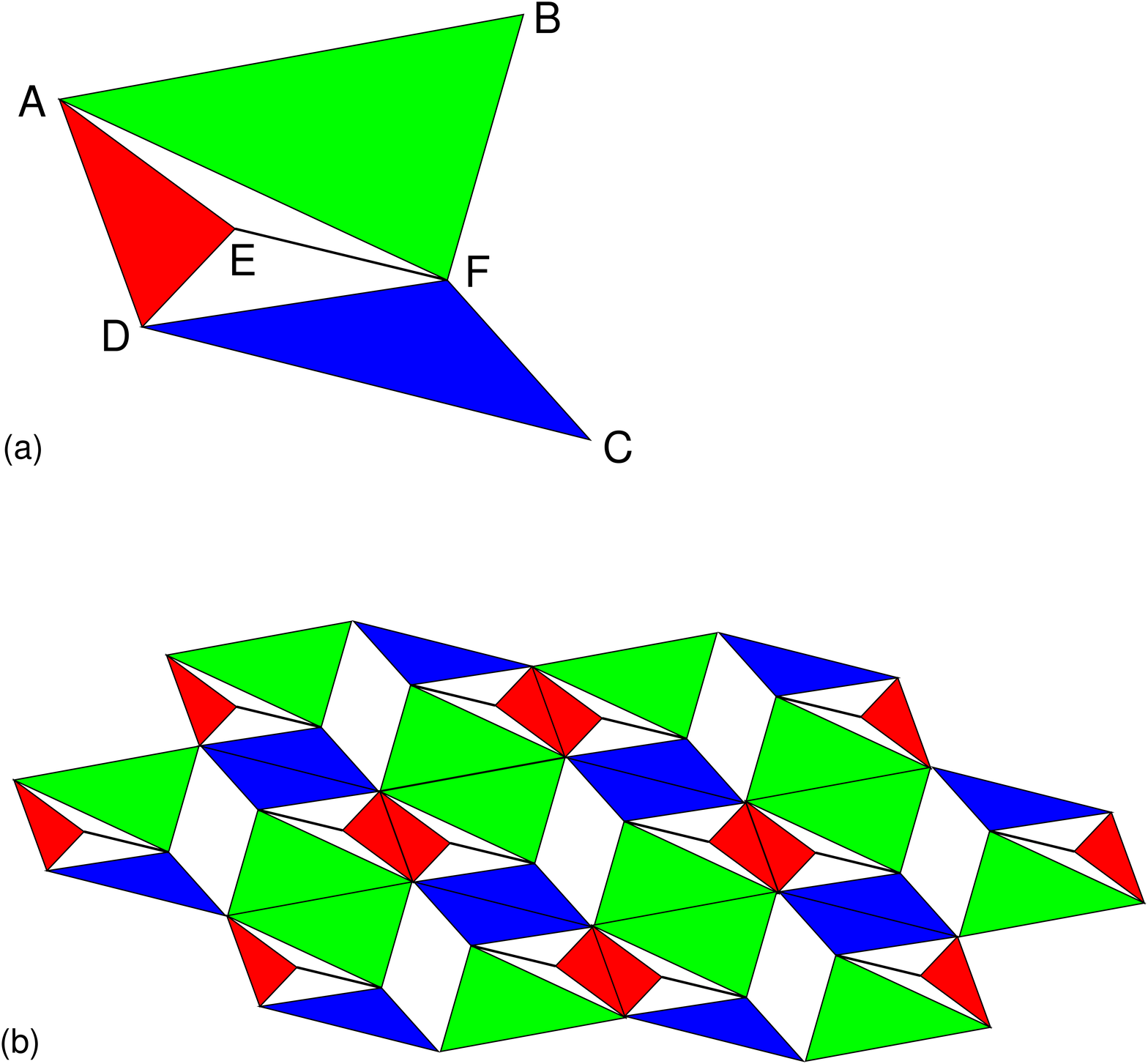}}}
\vspace{0.1in}
\caption{(a) shows a quadrilateral $ABCD$ which is a b-structure and which looks like the interior of (b) when tiled to fill all space. 
Here the blue, green and red triangles are u-structures, such as the one in figure \fig{2} and $EF$ is a rigid bar. 
Thus, for example, knowledge of the lengths $AD$ and $DC$, determines the triangles $ADE$ and $DFC$, 
and since the length $EF$ is fixed, the angle $ADC$ is determined, as is the length $AF$ and hence the triangle $ABF$.
Thus the quadrilateral $ABCD$ is determined (within a finite number of possibilities). Similarly if we know the
lengths $AB$ and $BC$, then given an angle $ABC$, we can determine the triangles $ABF$, $CDF$ (since $FC$ is known),
and $ADE$ (since $AD$ is known). That angle $ABC$ has to be chosen so $EF$ matches the length of the rigid bar, and then
the quadrilateral $ABCD$ is determined.}
\labfig{4}
\end{figure}

On the other hand, the structure of figure \fig{4}(a) is a b-structure. When it and its $180^\circ$ rotation are tiled to fill all space it looks
like the bimode material of figure \fig{4}(b). 
This bimode material, and similar bimode materials formed from 
other b-structures, have only two internal degrees of freedom.  To see this, consider the quadrilateral of \fig{5}(a), representing
the structure of figure \fig{4}(a), tiled to fill
all space. Now consider two adjoining tiles, as in \fig{5}(b), and suppose two actuators are added to cell 1, along say
the lines $BC$ and $AD$, and one actuator is added to cell 2, along say the line $EF$ slaved
so that tile 2 is always a $180$ rotation of tile 1, even as tile 1 deforms. No other actuators are added to the structure. 
If tiles 3 and 4 have the same shape and orientation as tiles
1 and 2 then they fit together perfectly as in figures  \fig{5}(c) and \fig{5}(d). Suppose however tile 3 deforms as in figure
\fig{5}(e). This changes the length $DG$ and the angle $CDG$. Tile 4 has to deform so the length $CD$ remains unchanged 
and the length $DH$ changes to the new value of the length $DG$. On the other hand, because tile 4 contains a b-structure the 
included angle $CDH$ will be determined (within a finite number of possibilities) by the lengths $CD$ and $DH$, and will not
in general equal the value of the new angle $CDG$: there will be an angle mismatch as in figure \fig{5}(f). To avoid this mismatch
the tiles 3 and 4 must deform as in \fig{5}(d). Similar considerations apply if we consider the deformation of the cells 5 and 6 and the cells 7 and 8 
adjoining this structure as in figure \fig{5}(g), and by induction for the deformation of the entire lattice of cells. Thus the deformation of the entire
lattice is controlled by the two actuators in cell 1 and the one actuator in cell 2 which is slaved. In fact we do not even need the
slave actuator in cell 2. If cell 2 deforms but cell 1 does not, then the deformation of cell pairs 3 and 4, 5 and 6, and 7 and 8
in figure \fig{5}(g) will be slaved to it, and in general there will be a misfit when we come to cell 9. 

This b-structure with its two internal actuators is an example of what we call a bimode transformer material: its macroscopic deformation is controlled by the two
actuators in the material, and lies along a two dimensional surface in the three dimensional space of invariants
describing the macroscopic deformation (excluding translations and rotations). From a practical viewpoint it would be better to have a sparse array
of slaved actuators rather than just two, to avoid enormous stresses on the actuators when the structure is loaded. Also if the bars are not completely rigid, the
desired deformation would not be transmitted to the entire structure from just two actuators, so again a sparse array would be better. 

\begin{figure}[htbp]
\vspace{6.0in}
\hspace{0.5in}
{\resizebox{2.0in}{1in}
{\includegraphics[0in,0in][7in,3.5in]{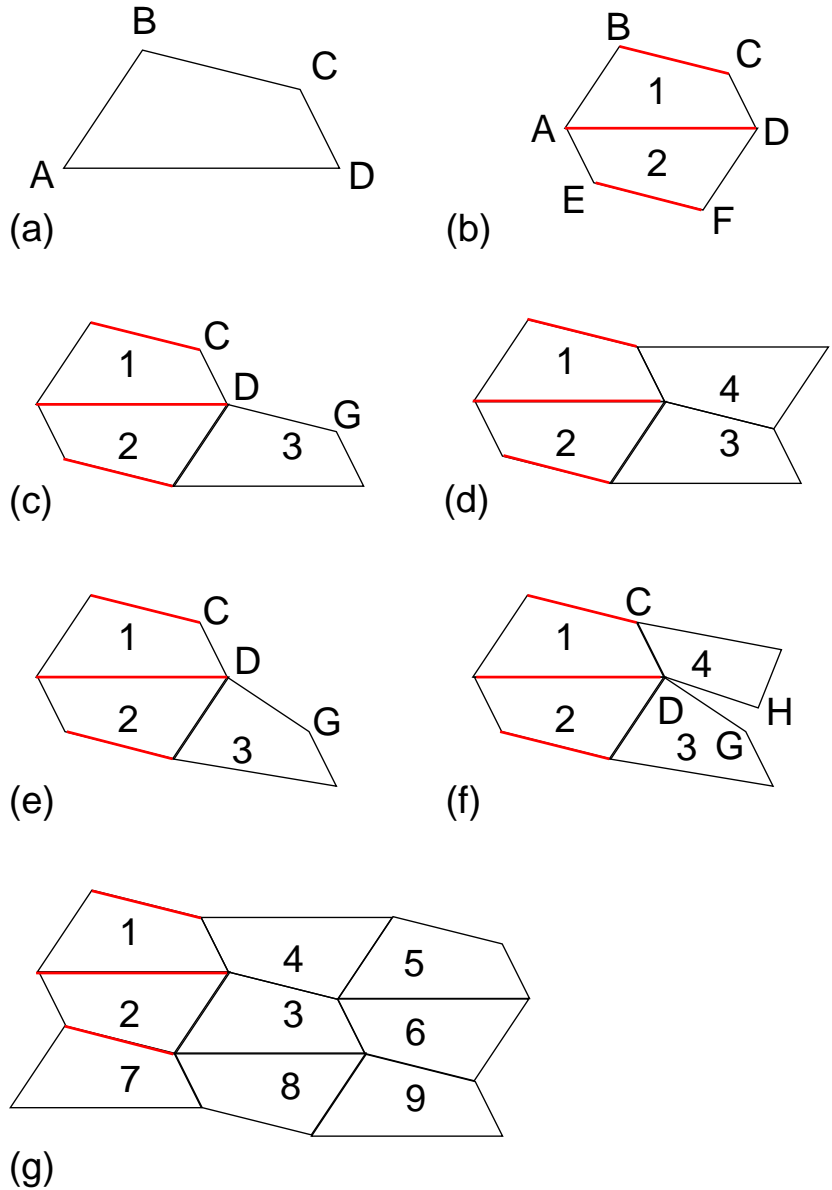}}}
\vspace{0.1in}
\caption{Sketches showing that the deformation of a tiling of b-structures is determined by the deformation of one tile. Actuators are positioned along
the red lines, and finally it is argued that the actuator along the line $EF$ is not needed. See the text for more explanation.}
\labfig{5}
\end{figure}

\section{On the route towards trimode transformer materials}
\setcounter{equation}{0} 
Ideally one would like to construct trimode materials whose deformation is controlled by actuators in just a few cells.
It is not clear if they exist or not, but one route towards their realization might be the construction of
what we call a t-structure (t for trimode). It is a hexagon $ABCDEF$ with opposite sides being parallel and of the same
length as shown in figure \fig{6}(a). The sides are not rigid, and there are three internal degrees of freedom. The triple
consisting of the lengths of any pair of adjoining sides, such as $FA$ and $AB$, plus the included angle can range any 
 anywhere inside some open connected region of three-dimensional space (i.e. the triple is not confined to lie on a surface)
and knowledge of the triplet completely specifies the hexagon  up to a finite number of possibilities. At this time we
do not know of a mechanism of bars and pivots which realizes t-structures, nor if such a mechanism even exists. 

Assuming t-structures can be made, we could consider a tiling of space with them. Now consider a cluster of two adjoining tiles, as in \fig{6}(b), 
and suppose there are six actuators, three in each cell, slaved
so that tile 2 is always a copy of tile 1, even as tile 1 deforms. Since the lengths $GC$ and $CD$ and the included angle $GCD$ are specified 
the adjoining tile 3 can only deform in synchrony with tiles 1 and tiles 2. By induction the same argument applies to any cell in the tiling:
the deformation of the entire structure is determined by the three actuators in cell 1, and their three slave actuators in cell 2.

\begin{figure}[htbp]
\vspace{3.0in}
\hspace{0.5in}
{\resizebox{2.0in}{1in}
{\includegraphics[0in,0in][7in,3.5in]{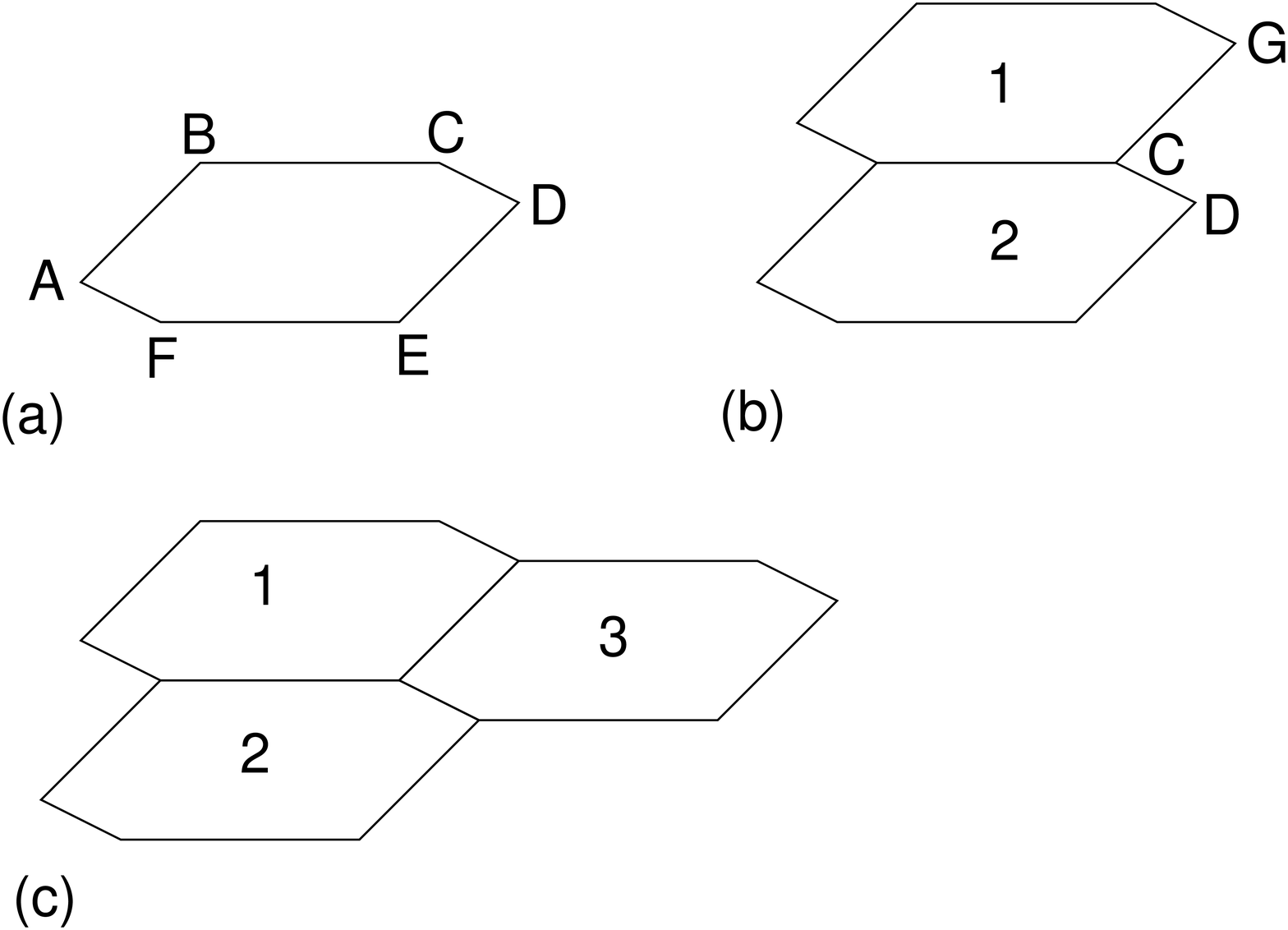}}}
\vspace{0.1in}
\caption{Sketches showing that the deformation of a tiling of t-structures is determined by the deformation of tile 1 assuming tile 2 is slaved to it.
See the text for more explanation.}
\labfig{6}
\end{figure}

Now consider what we call a trimode transformer material, a tiling of t-structures where actuator cells and their adjoining slave cells are sparsely populated
throughout the material, as illustrated in figure \fig{7}.  Also suppose the bars inside the t-structures are not completely rigid, but have some small flexibility. Then 
the deformation in the neighborhood of an actuator cell should be controlled by its three actuators. So if the settings of the actuators 
vary slowly from one actuator cell to its nearest neighbor actuator cells, then it seems likely (assuming there is no buckling) that we could control (within limits) the macroscopic inhomogeneous
deformation of the material. This may be very useful for practical applications. Of course one could do this just with a triangular array of actuator bars, but the 
goal is to reduce the number of actuators in the construction. 
 
\section{Open problems}
\setcounter{equation}{0} 

There are many open problems remaining. Within the class of non-affine metamaterials using rigid bars and pivots it would be interesting to know what macroscopic deformations (besides the affine
ones) can be realized by some structure. Although we have provided examples of a large class of bimode materials, it is doubtful that this class is rich enough to capture all types of 
affine macroscopic deformations that bimode materials can exhibit. It would be wonderful to have a characterization of possible macroscopic deformations analogous to that for affine unimode materials 
(\citeAY{Milton:2012:CCM}). As mentioned, it is an open question as to whether there exist t-structures as described in the previous section.  Even if t-structures do not exist, 
there may be other, yet undiscovered, routes towards realizing trimode transformer materials. Finally the territory of what affine and non-affine macroscopic deformations are possible
in three dimensional bimode, trimode, quadramode, pentamode and hexamode structures remains largely unexplored: we are even lacking examples of non-linear quadramode and pentmode materials.

\begin{figure}[htbp]
\vspace{2.0in}
\hspace{0.0in}
{\resizebox{0.5in}{0.25in}
{\includegraphics[0in,0in][10in,5in]{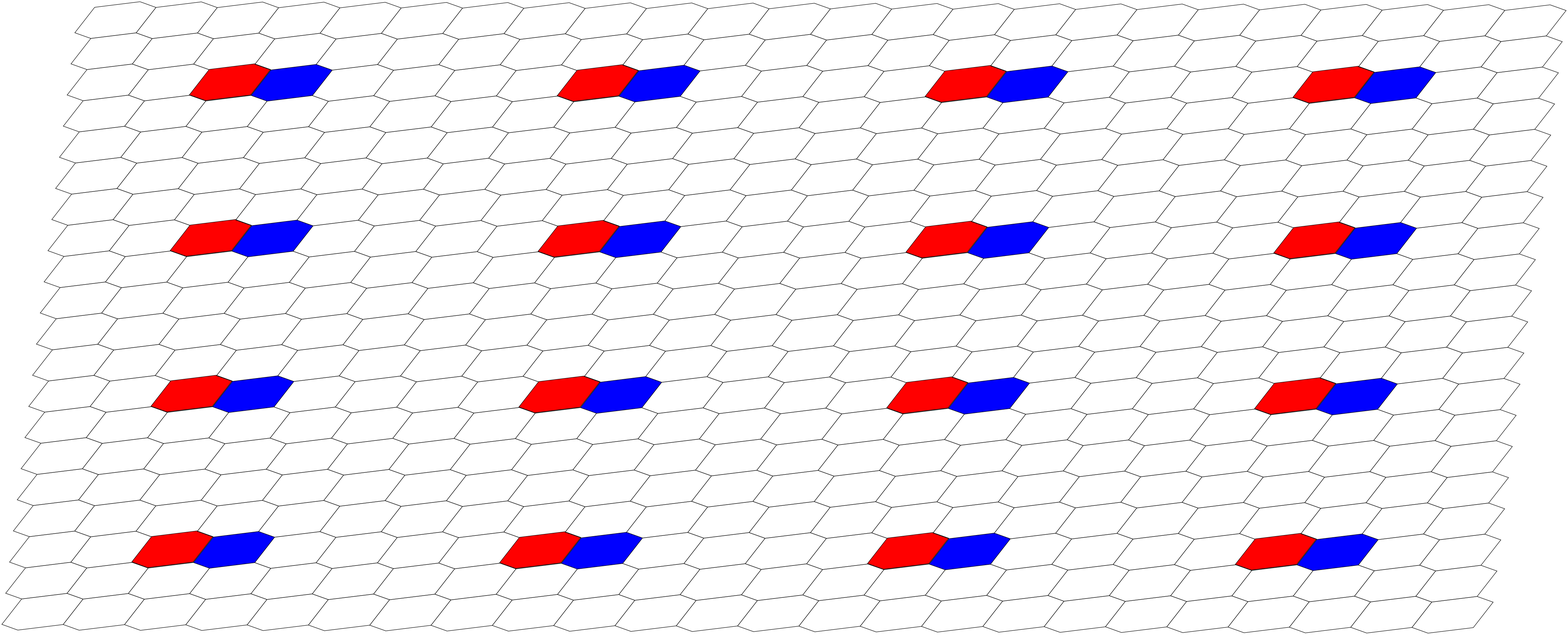}}}
\vspace{0.1in}
\caption{A tiling of t-structures containing a sparse array of actuator cells (red) and their adjoining slave cells (blue). }
\labfig{7}
\end{figure}

\section*{Acknowledgements}
GWM is grateful for support from the National Science Foundation through grant DMS-1211359.

\bibliographystyle{/u/ma/milton/tex/mod-xchicago}
\bibliography{/u/ma/milton/tcbook,/u/ma/milton/newref}
\end{document}